\documentstyle[psfig]{mn}

\newcommand{\alt}{\ \raisebox{-.3ex}{$\stackrel{<}{\scriptstyle \sim}$}\ }

\newcommand{\degree}{$^{\circ}$}

\newcommand{\mum}{$\mu$m}

\title[DENIS and the Inner Galaxy]{The Inner Galaxy resolved at IJK using
DENIS data}

\author[M.Unavane et al.]{
M.~Unavane,$^1$ Gerard~Gilmore,$^{1,4}$ N.~Epchtein,$^3$ G.~Simon,$^2$
D. Tiph\`ene,$^3$ B. de Batz,$^2$\\
$^1$ Institute of Astronomy, University of Cambridge,
Madingley Road, Cambridge CB3 0HA, UK
\\
$^2$ Observatoire de Paris-Meudon, DASGAL, CNRS/URA335, 5 place Jules Janssen,
F-92195 Meudon Cedex, France
\\
$^3$ Observatoire de Paris-Meudon, DESPA, CNRS/URA264, 5 place Jules Janssen,
F-92195 Meudon Cedex, France
\\
$^4$ Institut d'Astrophysique de Paris, 98bis Boulevard Arago,
F-75014 Paris, France
}

\date{\today}

\begin{document}
\maketitle
\begin{abstract}

We present the analysis of three colour optical/near-infrared images, in IJK,
taken for the DENIS project.  The region considered covers 17.4 deg$^2$ and lies
within $|l|<5$\degree, $|b|<1.5$\degree. The adopted methods for deriving
photometry and astrometry in these crowded images, together with an analysis of
the deficiencies nevertheless remaining, are presented. The numbers of objects
extracted  in I,J and K are 748\,000, 851\,000 and 659\,000 respectively, to
magnitude limits of 17,15 and 13. 80\% completeness levels typically fall at
magnitudes 16, 13 and 10  respectively, fainter by about 2 magnitudes than the
usual DENIS limits due to the crowded nature of these fields. A simple model
to describe the disk contribution to the number counts is constructed, and
parameters for the dust layer derived.
We find that a formal fit of parameters for the dust plane, from these data in
limited directions, gives a scalelength and scaleheight of 3.4$\pm$1.0\,kpc and
40$\pm$5\,pc respectively, and a solar position 14.0$\pm$2.5\,pc below the
plane. This latter value is likely to be affected by localised dust
asymmetries.
We convolve a detailed model of the systematic
and random errors in the photometry with a simple model of the Galactic disk and
dust distribution, to simulate expected colour-magnitude diagrams. These are in
good agreement with the observed diagrams, allowing us to isolate those stars
from the inner disk and bulge. After correcting for local dust-induced
asymmetries, we find evidence for longitude-dependent asymmetries in the distant
J and K sources, consistent with the general predictions of some Galactic bar
models. We consider complementary L-band observations in a second paper.

\end{abstract}
\section{Introduction}

The central kiloparsec of the Galaxy is dominated by an extremely dense stellar
cluster of unknown origin and history, and poorly known properties. It
is unknown if this cluster is a remnant of the core about which the
Galaxy grew, is the product of a later merger, is a product of a
long-lived bar in the disk feeding gas into continuing star formation
in the central galaxy, or has some other history.

Its relationship, if any, to the larger Galactic bulge, halo, and disk
and to the smaller Galactic non-thermal nucleus is entirely
unknown. This cluster, whose density approaches $10^6$ solar masses
per cubic pc, or $10^7$ times that of the Solar neighbourhood, is the
most extreme dynamical system available for detailed study.

The central degree or so is also an extended X-ray source, with
temperature some $10^8$ K, and gas pressures 1000 times those of the
normal ISM. Moreover, the cluster changes its luminosity density profile
by 2 in the power law index in some unobserved region between the
central few arcsec and the optically observable region some degrees
away.
How and where? And are more complex spatial distributions possible?
For example, in M31, the nearest similar spiral, the
central region shows two luminosity maxima, neither of which
corresponds to the centre of the larger scale gravitational potential,
or is understood. \cite{vdm97}

In practice, because of the high extinction, it is necessary to work
in the infrared.

Many high spatial resolution near IR wavelength studies of the central
arcminute or so are available (cf Genzel, Hollenbach \& Townes 1994) for
a detailed review). However, remarkably little data is available
concerning the larger scale structure. Balloon and satellite surveys
(e.g.  COBE/DIRBE) are all of very low spatial resolution. The only
large higher resolution survey, covering about $2^\circ \times
0.5^\circ$ around the galactic centre, is by Glass, Catchpole, and
Whitelock (1987).  It was performed in the J, H and K
bands up to a limiting magnitude of K=12.  While almost the entire J
map was dominated by heavy interstellar extinction, those at H and K
show progressively more detail of the inner region. They show clear
changes in spatial structure for different populations, suggesting
that analysis of low resolution data will necessarily be problematic

For the galactic centre this means that M and late K giants can be
reached in K but not at the shorter wavelengths as the extinction will
be too strong (up to 5 magnitudes in J [Catchpole, Whitelock \& Glass 1990]
leading to an expected apparent J magnitude of $\approx
16.5$~mag; [Wainscoat et al. 1992]). We also expect that
essentially all I and most J objects seen in the plane will be disk objects.

Two large scale, high resolution surveys in near infra-red bands have
begun recently. 2MASS (2-micron all sky survey)  \cite{skr97} aims to survey the whole celestial sphere in J,H and
K, with 2 arcsecond pixels, from two specially built identical
telescopes in each hemisphere.

The second project is a European joint venture, called DENIS (Deep
Near Infra Red Southern Sky Survey) which aims to map the whole of
the southern sky in I,J and K with 3 arcsecond pixels. (Epchtein et al
1997). Technical details are given in Copet et al. (1997).

It is the data from this project which are relevant to the central
regions of the galaxy that we consider in detail here, considering specifically
the ability of DENIS-like survey data to study the inner Galaxy.

\section{DENIS -- data reduction}

\begin{figure}
\label{dr}
\end{figure}

\subsection{Using the images}
DENIS (DEep Near InfraRed Southern Sky Survey)
(Epchtein 1997) will
be a complete deep near infrared survey of the southern sky, with the
objective to provide full coverage in 2 near infrared bands (J at
1.25$\mu m$ and K at 2.2$\mu m$) and one optical band (I at 0.8$\mu
m$), using a ground-based telescope and digital array detectors.
Spare time at the end of DENIS observation nights during
summer/autumn 1996 were used to take images in rasters (see Figure
\ref{dr}). Note that the shaded circle in this figure indicates the size
of the DIRBE beam for comparison.

The standard pipeline processing of images ensures that for each
image, sky subtraction is performed, the background is made flat, so
that the images are ready for extraction of sources. (Borsenberger, 1997).
 The early pipeline extraction procedures proved unsatisfactory for
these very crowded fields.

\subsection{Source Extraction}

Source extraction and aperture photometry were performed with the
SExtractor software
\cite{sex}. Sources were extracted to 2$\sigma$ above the background
noise level, and and aperture of 3.5 arcseconds radius was
used. The calibration and flat fielding
carried out in the preprocessing of
the images is assumed to be sound. It is important to note that if a
background map were to be constructed for each image as part of
the source extraction procedure, it would remove
one of the effects being studied (namely, the small scale variation of
the extinction).

The most crowded fields have $\sim$ 8000 sources in K, corresponding
to a mean separation of $\sim$ 10 arcsec, some 3 times larger than the
radius of the aperture. The zero point was taken from standard star
frames taken immediately before and after the observation of each
raster.

\subsection{Source matching between IJK}

Each frame suffers its own distortions -- software was developed to
take triplets of corresponding I,J and K catalogues for a single image
field, and map all onto a single consistent coordinate system. The
chosen coordinate system, to make mapping straightforward, was chosen
to be the coordinate system of the J frame. The choice is
astrophysically motivated. The difference in the nature of the
brightest sources in the presence of high extinction varies so much
between I and K that matching these two would be unreliable if
automated. The technique adopted uses the iterative fitting, with rejection
of outliers, of a 2nd order, two-dimensional polynomial to several hundred of
the brightest sources for the
transformation between images. Higher-order transformations are found to be
no better.

\subsection{Absolute astrometry}

The usefulness of these data cannot be fully realised unless cross
matching with other databases is possible. To this end, {\it absolute}
astrometry is required. After transformation of the positions of the I
and K images to the J reference frame, we need to find a
transformation between this J frame, and an absolute $(\alpha,\delta)$
coordinate system. Examining available catalogues, we find that as of
the time of writing, the catalogue with the greatest number of
astrometric objects is the Hubble Guide Star Catalogue (GSC). This
catalogue contains nearly 19 million objects, designed to satisfy the
operational needs of the Hubble Space telescope. The objects are thus
distributed almost evenly across the sky.

A DENIS image is about 12 arcminutes on a side, which means that, on
average, the number of GSC sources found in an image is 18. However,
the coverage is not entirely homogeneous. Near the Galactic plane, the
regions of particular interest in this case, there are far fewer
sources. For example, the  $1^{\circ} \times 1^{\circ}$ around
$\ell=0^\circ,b=0^\circ$ contains only 307 GSC sources, which corresponds, on
average, to 12 per DENIS image. Poisson statistics tell us that there
is then a 5\% chance of $\le 6$ objects being found in an image. In
fact, the patchiness of foreground extinction in these regions means
that even Poisson statistics are an unreliable guide as to the number
of objects we may find per frame.

Furthermore, at least second order fits are likely to be necessary to
derive positions based on astrometric points, simply judging by the
fact that a 2nd order transformation was needed between I and J images
(taken by the same instrument, but using different optical paths, and
detectors.)  An average of 12 points per frame is not good enough to
fit the 12 coefficients required for such a fit -- on average, the
solution will be numerically unstable by being forced to provide 12
coefficients from just about 12 points, and about half the time (at
least), there would not be enough points at all.

\subsection{Digitised Sky Survey}

The Digitised Sky Survey (DSS) is available on a set of CD-ROMS, and
also online. In order to produce a list of astrometric points for use
in performing absolute astrometry with DENIS, a viable option is to
load relevant regions of the Digitised Sky Survey, source extract the
images, and fit the GSC stars in that image to provide an astrometric
database with a far greater density of objects.

For the region of interest indicated in fig \ref{dr}, a grid of
$1^{\circ} \times 1^{\circ}$ DSS images was overlaid to cover it. Each
image was source extracted using SExtractor, the acceptance criterion
being $2\sigma$ above background. The aperture `magnitudes' obtained
are unimportant, and do not relate linearly to the true magnitudes,
since these images derive from photographic plates. However,
they are useful in limiting the number of sources used for matching to
the GSC, to a few hundred.

The GSC typically has 300 stars deg$^{-2}$ in these regions. As in the
case of the IJK images, the same software was used to make the
transformations between $(x,y)$ in the image and
$(\alpha,\delta)$ as
determined by the GSC stars, projected to the local tangent plane.

 The increase in density
of astrometric reference points results in a density
about 50 times greater than when using the GSC alone, leading
to, on average, 500 points per DENIS image. We call this our
auxiliary
astrometric catalogue (AAC).

\subsection{Using AAC}

The I band images, taken at $\sim$0.8\mum, correspond closely in
wavelength to the scans used for the DSS, and matching sources between
them is a relatively straightforward exercise. Using the same
technique  as before, a 2nd order, two dimensional polynomial fit is
made between the I images and the AAC extracted from the DSS.
The catalogue of I images used is already transformed
to J-image coordinates, so that the resulting
transformation is directly from J $\rightarrow$ AAC. Finally this J
$\rightarrow$ AAC transform can be applied to all the image source
coordinates to give 3 separate catalogues in I,J and K.

\section{Final catalogue}

For each of the I,J and K bands, and for each raster as shown in
figure \ref{dr}, a final one-band catalogue is made, and sources
duplicated in the overlaps of frames are removed. The overlapping
images are used to give estimates of random astrometric and
photometric scatter. (See later section on random errors)

For each raster, these large I,J and K catalogues are matched
up. Seven classes of source are distinguished -- those present in IJK,
those present in JK only, those in IJ only, those in IK only (rare!),
those only in I, those only in J and those only in K. The matching is
carried out by assigning initially as the same,
objects closer than 3 arcseconds
to one another in different bands.

The output catalogue contains Right Ascension and Declination {\em for
each of the 3 bands separately}, a symbol indicating which of the 7
classes the object belongs to, and I and/or J and/or K magnitudes. The
reason for keeping all 3 positions is explained in the following
sections.

\subsection{Astrometric and photometric precision of final catalogues}

\subsubsection{Random errors}

\begin{figure}
\label{astromerr1}
\end{figure}

\begin{figure}
\label{photerr}
\end{figure}

The separations between matched objects in a complete raster (in this
case C02), in each colour catalogue separately, is shown in figure
\ref{astromerr1}. Some 70\%, 80\% and 75\% respectively, in the I,J
and K catalogues, have the same objects within 1 arcsecond of one
another (no mean feat for 3 arcsecond pixels!).  The random
photometric scatter can be seen in the left of figure \ref{photerr}, again by
comparison of duplicated sources from image overlaps. The figure
indicates the difference between magnitudes $m_1$ and $m_2$, measured
from different images. A fit to the core values every $\frac{1}{2}$
magnitude gives values for the standard deviation of the difference
$m_1 - m_2$, and since $\sigma^{2}(m_1 -
m_2)=\sigma^{2}(m_1)+\sigma^{2}(m_2)$, $m_1$ and $m_2$ coming from the
same underlying population, the standard random error in one
measurement of the magnitude is given in each case by the $\sigma =
\sigma(m_1 - m_2)/\surd 2$. This distribution is indicated in
the right panel of figure
\ref{photerr}.

\subsection{Systematic errors}

The major systematic errors in magnitude measurement will be the
result of crowding in the fields. The pixelsize used in the DENIS
cameras is 3 arcseconds inevitably making it difficult to properly
resolve sources in the crowded galactic centre regions.

We make an estimate of the size of this systematic photometric
uncertainty by adding artificial stars to DENIS images. The images are
then processed just as for the unaltered images, and the extracted
magnitudes are compared to the input magnitudes.

These simulated sources are analgous to individual real sources -- they can be
considered as single additional sources in the presence of very many nearby
sources. The statistics of the difference between the observed and generated
magnitudes will be representative of the same statistics for real sources. (e.g.
see Sodemann \& Thomson 1997)

A crowded image (at roughly $\ell=$0\degree\ and b=1\degree) was used as the
base onto which artificial stars were added. Average values for the
parameters of full-width at half-maximum, ellipticity, and orientation
of the extracted sources in that image were found, and these were used
to generate random sources. The sources were distributed at random
over the image in steps of 1 magnitude from 11 to 17 in I, from 9 to
15 in J, and from 7 to 13 in K. Each image had 225 stars added (this
is a small fraction when the number of sources per image is typically
4--5000). The resulting 7 images in each band were treated as described
above for the untreated images.

The derived catalogues were searched for the artificial stars to a
distance of up to 3 pixels from their input positions. If found, their
magnitudes and positions were noted. The magnitudes were characterised
by a mean offset and a scatter about that mean offset.

The distribution of magnitude offsets is shown in figure
\ref{syserr}. At the brightest magnitudes, there is little systematic
shift in the magnitude due to the source extraction procedure. As
higher magnitudes are reached, the measured magnitude (calculated
using an aperture of 7 pixels diameter as indicated above) is
systematically brighter than the actual magnitude. This can be
understood in the context of the severe crowding as the result of the
flux of nearby sources entering the aperture.

\begin{figure}
\label{syserr}
\end{figure}

\subsection{Completeness}

Using the same simulations as above, the numbers of sources recovered
within a given tolerance of the nominal position can be assessed and
used as a measure of the completeness in these crowded fields. Figure
\ref{completeness} shows the fraction of simulated sources recovered
from a typical crowded DENIS image, for various tolerances from the
nominal position (1$"$, 2$"$ and 3$"$).

\medskip

\begin{tabular}{cccc}

Deviation (arcsec) & \multicolumn{3}{c}{80\% Completeness} \\
 & I & J & K \\
\\
1 & 15.8 & 12.6 & 9.9 \\
2 & 17.0 & 13.6 & 11.3 \\
3 & $>$17 & 13.9 & 12.1 \\
\\
\end{tabular}

Note that these values are inferior to the expected limits for the DENIS
survey in general, since we are confusion limited in very dense fields.

\begin{figure}
\label{completeness}
\end{figure}

\subsection{Multiband completeness}

For ease of data treatment, the colour magnitude diagrams are made by
considering only sources matched between pairs of images less than a fixed
distance apart. In the next section, for the analyses, we use 1
arcsecond. This value has no meaning when considering the statistics
only of I, J or K data, since we are not interested in the absolute
precision with which the source position has been found, but only with
the numbers of such sources at given magnitudes.

However, a tolerance of 1 arcsecond (or some other value) plays an
important role in determining the distributions in IJ, JK and IJK
statistics.

Presented below are the results of a Monte-Carlo technique for
determining what fraction of IJ,JK and IJK matched images will be
found if an upper limit is placed on their separation. The same
artificial star experiments as above are used. We use the distribution
of displacement from nominal position with magnitude (as indicated in
a discrete manner in figure \ref{completeness} for three values of
displacement).

Taking the case of J and K as an example, we use the
probability distribution for J and K for the displacement from the
nominal position and build up a 2-dimensional grid  for the fraction
retrieved according to the stringent criterion that the J and K
sources lie within a given distance of one another.

Points are generated according to the measured probability distribution function, and the
acceptance fraction is deduced using the numbers of J and K points
which lie within 1 arcsecond of each other. The contour plot in figure
\ref{jkcontour} shows the result of this calculation, plotted on the
same J$-$K,K plane as for the other plots.  Shown in the figure are
contours showing the probability of retrieval of a J$-$K pair, if the
relative displacement tolerance is set at 1 arcsec. As expected, for
objects bright in both bands (ie. low K and low colour), the retrieval
probability is high, and this probability falls markedly as the K
magnitude and/or colour is increased.

\begin{figure}
\label{jkcontour}
\end{figure}

This diagram, smoothed by the fitting of a low-order two-dimensional
polynomial is used below as a convolving mask applied to model
colour magnitude diagrams to enable a quantitative comparison with the
observed diagrams.

In exactly the same manner, an analogous analysis is carried out for the case
of the I$-$J and J colour magnitude diagram. The same features are seen (bright
blue objects are almost all retrieved, while dimmer, redder objects are not so
well retrieved).

Finally, in the case of the two colour I$-$J and J$-$K diagram,
contributions to a given part of the diagram come from objects of
different magnitudes. At the
brightest K magnitudes, as expected, the regions of low J$-$K value
are essentially complete, and as we go to fainter K magnitudes, the
completeness gets progressively lower. In a similar way, a three dimensional
(K,I$-$J and J$-$K being the variables) polynomial is used as a mask.

\subsection{Comments about completeness levels assigned}

The test image used to derive these completeness results was an image
at latitude $\sim$1\degree, where the source density in I,J and K is the
highest of any position in the fields observed. In this sense, the
completeness levels and estimates of photometric shifts are worst-case
estimates. In general, completeness will be better, and photometric
shifts less.

\section{Looking for structure}

\subsection{Overview}

The data reduction described in the previous sections results in a
wealth  of point source data in the I,J and K bands. The region
observed forms an irregular polygon contained within $|l|<$5\degree\ and
$|b|<$1.5\degree.  The latitude coverage is greatest nearest  to zero
longitude, just as the longitude coverage is greatest near zero
latitude. Figure \ref{dr} shows the coverage. The data is severely
confused due to the large  (3 arcsecond) pixels used coupled with the
huge numbers of sources at low  latitudes. The effects of crowding on
photometry and completeness are described above.

In this section, we first present colour-magnitude diagrams for
different parts of the region covered, and attempt to gain a
qualitative understanding. Many of the unusual morphological features
in the CM diagrams can be understood in terms of crowding and
completeness problems. The major effect seen in these diagrams is
clearly that of a prominent, thin dust lane running through the fields
centred near b=0.

A simple model is constructed to understand the colour magnitude
diagrams resulting from observations of sources in a stellar disk in
the presence of a strong dust layer.

With the aid of latitude-colour diagrams, we establish, using this
model, parameters to describe the dust layer, and compare synthetic
colour magnitude  diagrams thus derived to the observed ones. It
becomes clear that disk stars alone can describe many of the features
in the appearance of the CM diagrams.

This model is then used to deduce the distribution in distance of the
sources seen, in a statistical way, from the distribution in magnitude
that results from the observations. Magnitude cuts are made in the
luminosity functions corresponding to two regions, one dominated by
near-disk objects, and the other by far-disk/bulge objects.

As a test of various Galactic bar models, the cuts
above are used to remove the effect of nearby disk asymmetries, and to
penetrate the far disk and central bulge regions. In J and K,
asymmetries in the same sense as predicted by bar models are seen at
the 3$\sigma$ level.

\subsection{Looking at the data}

The dataset derived in the previous section is in three passbands
(DENIS I,J and K centred at 0.8\mum, 1.25\mum\ and 2.2\mum). The number
of images processed is 613 in each passband, each covering an area of
770x770 arcseconds. The area covered is  about 17.44 deg$^2$, which
means that double coverage  occurs for 10.60 deg$^2$ (or about
60\%). Excluding the poor quality data very near the edges, this
becomes roughly 50\% overlap.  It is this which has allowed a good
characterisation of random photometric uncertainties.

\subsection{Numbers}

The total number of sources extracted is some 1.500$\times$10$^6$ in
I,  1.707$\times$10$^6$ in J, and 1.324$\times$10$^6$ in K. After
removal of duplicate observations the numbers become
0.748$\times$10$^6$ in I,  0.851$\times$10$^6$ in J, and
0.659$\times$10$^6$ in K.

\subsection{Colour-Magnitude and two-colour Diagrams}

The following diagrams, (figures \ref{blt0.1} -- \ref{c02} and
later) , indicate, in the bottom left hand corner, a two-colour
diagram with (J$-$K) on the abcissa and (I$-$J) on the ordinate. The
top left panel in each case shows the colour magnitude diagram of
J$-$K against K, with the (J$-$K) scale of the lower diagram being
preserved. Finally, the bottom right panel shows the (I$-$J) against
J colour-magnitude diagram  rotated anticlockwise by 90\degree\ to
match the (I$-$J) scale in the two-colour diagram. This is indicated
diagramatically in the explanatory figure \ref{explain}. Note that
each diagram is constructed separately, so that the number of
points in the two-colour diagram will be fewest of all (requiring
coincident I,J and K sources).

\begin{figure}
\label{explain}
\end{figure}

The first three sets of diagrams are for 0.2\degree\ latitude cuts
between $-$0.1\degree\ and 0.1\degree, 0.4\degree and 0.5\degree\, and 1.0\degree\
and 1.2\degree\.  (Figures \ref{blt0.1}, \ref{b0.4-0.6},
\ref{b1.0-1.2}). On each diagram is also marked a reddening line,
calculated as in a later section.

Also shown, in figure \ref{c02}, is a cut for a region around $\ell=$3\degree\ and
b=0\degree.

\begin{figure}
\label{blt0.1}
\end{figure}

\begin{figure}
\label{b0.4-0.6}
\end{figure}

\begin{figure}
\label{b1.0-1.2}
\end{figure}

\begin{figure}
\label{c02}
\end{figure}

Before commenting on these digrams, it is worthwhile to present
colour-magnitude and colour-colour diagrams  in the bands I,J and K
for disk type III and type V objects. These will enable a comparison
to be made. Figure \ref{cmstandard} shows these diagrams. The data for
the absolute K magnitudes are taken from a study by Garwood \& Jones
(1987), who produced a local disk luminosity function. The
transformations to I and J are made according to data tabulated by
Zombeck (1990). Notice that despite the wide range in absolute
magnitudes, both the I$-$J and J$-$K colours have a very small intrinsic
spread (\alt 1).

\begin{figure}
\label{cmstandard}
\end{figure}

The magnitude cuts in I,J and K respectively are at magnitudes 11,9 and 8 for
the bright end (there are saturation problems brighter than this) and at 17,15
and 13 at the faint end.

The most marked feature in figures \ref{blt0.1},\ref{b0.4-0.6} and
\ref{b1.0-1.2} is the straight line seen in the two-colour diagrams,
which follows the reddening line. This demonstrates more clearly than
anything else how important a role interstellar extinction plays in
the interpretation of these diagrams.  From figure \ref{cmstandard},
it is clear that intrinsic stellar properties will not contribute more
than a magnitude of colour shift, so these diagrams immediately
indicate that extinction in the line of sight of at least $A_V$=10--15
is to be expected. These figures also show some scatter to the right
of the reddening line. This can be accounted for by crowding effects
as will be demonstrated in a later section. Also, in the two-colour
panel in figure \ref{b1.0-1.2}, the highest latitude field, there is a
clear bend at the reddest part of the line, so that objects are redder
in (I$-$J), or equivalently, bluer in (J$-$K) than expected. Again, this
is an effect which can be understood for this dataset in terms of
crowding problems, and is discussed later.

The other panels in these figures are less easy to interpret. Some
morphological features are clear. In the (J),(I$-$J) panels, there are
clear striations running almost parallel to the reddening line. The
very dominant blue faint part of the diagram, clear at the lowest
latitudes, is much diminished at latitude $\sim$ 1\degree, while the
fainter redder part becomes dominant. This can be plausibly understood
in terms of a young, blue, main-sequence population whose presence
will be stronger at the lowest latitudes in the disk. Features which
are `reddened off the page' become progressively more apparent at
higher latitudes when the optical path through dust is lessened.

Similar features are evident in the (K),(J$-$K) diagrams, with the clear
concentration centred at a colour of J$-$K=4 in figure \ref{blt0.1}
being shifted to J$-$K$\sim$3.5 and J$-$K$\sim$2.2 in figures
\ref{b0.4-0.6} and \ref{b1.0-1.2} respectively. Again, this is
plausibly due to a line of sight which sees less of the absorbing
material in the disk.

Finally, figure \ref{c02} is included as an example of the
inhomogeneous distribution of absorbing material in the line of sight. In all
three panels, a clear break is seen in one of the striations. In the (J),(I$-$J)
panel, the striation starting at (J,I$-$J)=(13,3.6) and continuing to (11,2.0),
shows a clear break in the direction of the reddening line, corresponding to a
wall of extinction of $A_V \sim$ 3. The same break is seen in the (K),(J$-$K)
diagram in the striation starting at (J$-$K,K)=(2.0,12) and continuing to
(1.2,10). This too is reproduced in the next sections by means of a model.

\section{Simple disk model}

The approach we adopt in order to understand these colour-magnitude data is to
construct a model of the Galactic disk
from which we can generate sources, and make statistical
comparisons to the observed data. Details of this model,
and justification for its simplicity,
 are given in the appendix.
Only a summary is given here.
 We employ a model which
consists of a disk, exponential in both the vertical $z$-coordinate, and the
radial coordinate. The dust is also represented as an exponential
disk with its own
scaleheight and radial scalelength.

For the stellar disk, we use recently derived values for the
scalelength and scaleheight of 2.7\,kpc and 0.20\,kpc. (Freudenreich 1996,
Kent, Dame \& Fazio 1991)
 The small region of interest ($|l|<$5\degree, $|b|<$1.5\degree) is
not very sensitive to changes in these disk parameters, but rather
more to the dust layer parameters. The dust layer, and its appearance
to us, is characterised by four parameters -- the scaleheight of the
dust ($z_d$), the radial scalelength ($r_d$),
the height of the sun above the plane of this dust
($z_0$), and the strength of extinction in the midplane of the disk
($A_V$ in mag/kpc).  These are derived from a fit to the data.

The model output is convolved with the completeness levels and
photometric scatter derived above and is subsequently compared
to the observations. A sketch of the model geometry is given in figure \ref{sketch}

\begin{figure}
\label{sketch}
\end{figure}

\subsection{Luminosity Function}

The brightest sources visible at near-IR wavelengths, as
indicated in figure \ref{cmstandard}, are the later type giant stars.

\begin{table*}
%\caption{Transformation polynomials based on Zombeck (1990).}
\label{lumfntable}

%\begin{tabular}{ccccccccc}

%\multicolumn{9}{c}{The transformation from $m_K$ to $m_I$ and $m_J$ for disk giant stars.} \\

% & & $m_I$= & \multicolumn{5}{c}{0.7586 +0.7506$m_K$ -0.010689 $m_K^{2}$} & \\
% & & $m_J$= & \multicolumn{5}{c}{0.4497 +0.8262$m_K$ -0.011756 $m_K^{2}$} & \\

%\multicolumn{9}{c}{The transformation from $m_K$ to $m_I$ and $m_J$ for disk main sequence stars.} \\

% & & $m_I$= & \multicolumn{6}{c}{-0.179 +1.1896$m_K$ +0.0251 $m_K^2$ -0.00202$m_K^3$}  \\
% & & $m_J$= & \multicolumn{6}{c}{-0.076 +1.1005$m_K$ +0.0225 $m_K^2$ -0.00240$m_K^3$}  \\

%\end{tabular}

\end{table*}

\subsubsection{Giants -- Type III}

As a source for a luminosity function for the type III (giant) stars,
we refer to Garwood \& Jones (1987) , who observationally determined a local
luminosity function in the K-band. Colours for local objects are taken
from Zombeck (1990), who collates infra-red field star colours from a
variety of sources. Colours for each of the types of source are given,
from which a polynomial fit is made to determine an analytical
transformation from the absolute magnitude in K, $m_K$, to the
absolute magnitudes in I and J. ($m_I$ and $m_J$). A second
order polynomial was found to be sufficient in these case, and the
greatest  discrepancy with the tabular values was less than 0.1
mag. The polynomials used are tabulated in table \ref{lumfntable}.

\subsubsection{Main Sequence -- Type V}

Main sequence stars have intrinsically low near IR luminosities,
mainly due to their small size. What some main sequence stars lack in
size, they make up for in surface luminosity (i.e. high temperature)
so that the brightest of these stars (O and B) despite having their
peak in emission far from the near IR wavelengths we are interested
in, nevertheless show bright magnitudes at these wavelengths.

We adopt the K-band luminosity function described by
Garwood \& Jones (1987). Again, we
fit a low-order polynomial to fit the colours (J$-$K) and (I$-$K)
as given by \cite{zom}. A third order polynomial is sufficient to
prevent errors of greater than 0.1 mag. The polynomials are also shown
in table \ref{lumfntable}.

It is assumed in this model that the luminosity function $\phi(m_K)$
and the geometrical parameters $\rho(x,y,z)$ are independent. This is
an oversimplification since it is well known that the scale height of
stars in the disk depends upon the stellar type \cite{schm63}. Late
type V stars (hence typically older) are to be found with large
scaleheights while early type (and hence younger) stars have
distributions with smaller scaleheights. The mechanism is clearly
diffusive, as it is thought that stars form in the disk of the galaxy,
and over the course of several revolutions diffuse to distributions
with larger scaleheights. Giant stars, on the other hand, show an
essentially constant scale height with type.

However, the majority of sources seen at the near IR wavelengths are
the giant  stars and the older type V stars, which can all be taken to
have a (relatively) large scaleheight (some 2--300\,pc).

As a test of this, below we show the luminosity functions at various heights
above the plane, normalised to a scale height of 200\,pc, generated by using the
above luminosity functions, and scaleheight parameters for the different
stellar types. The biggest differences occur between absolute magnitudes 0 and 3
(depending on waveband) where there are few sources that feature in the
simulations presented. The three lines shown show the luminosity function
at 3 different heights in the plane -- z=0\,pc, z=200\,pc and z=400\,pc
(zero, one and tw
scaleheights of the oldest population).

\begin{figure}
\label{generate}
\end{figure}

We adopt a luminosity function for zero height (i.e. scale heights the
same for the different stellar types) for the reasons given above.

\subsection{Extinction coefficients for the DENIS filters}

Since large values of extinctions along the line of sight play a major
role in determining the appearance of the colour magnitude diagrams in
the plane of the galaxy, it is important to establish precise values
for the extinction coefficients.
We use the tabulation by Mathis (1990) for the relative extinction in
magnitudes as a function of wavelength. Between 1.25 \mum\ and 3.4 \mum, we
parametrise his tabulation as

$ \frac{A(\lambda)}{A(J)} = 1.484 - 5.60109 x +8.395624 x^2 -4.5947083 x^3 $

where $x=log_{10}(\lambda/\mu m)$. The error in this least-squares 3rd
order fit is not more than 4\%.

Referring to Copet et al. (1997), we use the
instrument+sky response profiles in each of the I,J and K bands, and convolve
them with the Mathis data to give the following values for
$A(X)/A(1.25\mu m)$:

\medskip

\begin{tabular}{ccc}
 Band (X) & $A(X)/A(1.25\mu m)$ & $A(X)/A(V)$ \\
  I &  1.968 & 0.554 \\
  J  & 0.994 & 0.280 \\
  K  & 0.396 & 0.112 \\
\end{tabular}

\medskip

For comparison, the Mathis (1990) values of $A(\lambda)/A_V$  for wavelengths
0.90$\mu$m, 1.25$\mu$m and 2.2$\mu$m, corresponding to Johnson's I,J
and K, are 0.479, 0.282 and 0.108.

\subsubsection{Application to model}

The chosen functional form for the dust layer is that of a disk exponential in
both radial and vertical components. The amount of the dust is represented in
terms of its absorption as:

$X(r,z) = \propto e ^{-z/z_d - r/r_d}$

where X is a measurement of $A_V$ in mag/kpc, and $r$ and $z$ are
galactocentric cylindrical coordinates. The function is normalized by setting
$dX/d\rho = X_0$ , where $\rho$ is distance along the line of sight and $X_0$ is
the local extinction per unit distance. The
functional form for the dust is integrated along the given line of sight to
yield a function $A_V(\rho)$ which represents the amount of extinction in the line
of sight up to a distance $\rho$ from the sun.

\subsection{Summary of model}

In summary, the model consists of the following steps. Details of the
Monte-Carlo method used to generate stars are given in an appendix.

\begin{enumerate}

\item For given lines of sight, a probability distribution function of distance
      is generated. An extinction/distance curve is also generated by
      integrating the
      function representative of dust along the line of sight.

\item By combining functions in the three wavebands with
      the distance distribution, a limiting
      parameter $w_0$ is found to speed up the Monte Carlo process (see
      Appendix).

\item Observable distance/stellar class pairs are generated, and K
      magnitudes are calculated from the luminosity function.

\item Analytical representations in terms of the K magnitude are
      used to give the corresponding I and J
      magnitudes

\item  To each absolute magnitude, the modified distance modulus is added
       ( 5$\lg$(r/kpc)+10+$A_X$, where X=I,J or K).

\item To each magnitude, is further added a random scatter in magnitude as
   determined by the artificial star experiments above.

\item A systematic magnitude offset to simulate crowding/extraction effects is added.

\item The source is rejected if its magnitude falls outside specified limits.

\item Steps 3--8 are repeated until the specified numbers of sources have been generated.

\end{enumerate}
\section{Model output}

Figures \ref{model0.0}, \ref{model0.5} and \ref{model1.1} show the
results obtained from running the model for three different latitudes
b=0\degree, b=0.5\degree, and b=1.1\degree, which may be compared with
figures \ref{blt0.1}, \ref{b0.4-0.6} and \ref{b1.0-1.2}
respectively. Figure \ref{modelwall} shows the result of a simulation where a
cloud of extincting material with $A_V$=3 has been placed between 1.5 and 2.5\,kpc from the sun in the line of sight.
The raw output from the model has been treated with the
systematic shifts and random scatter in photometry due to crowding,
and the colour-magnitude and colour-colour diagrams have been
convolved with the completeness masks derived at the end of the first
section.

\begin{figure}
\label{model0.0}
\end{figure}

\begin{figure}
\label{model0.5}
\end{figure}

\begin{figure}
\label{model1.1}
\end{figure}

\begin{figure}
\label{modelwall}
\end{figure}

We stress again that these models are constructed for a disk only, and
it is expected that whatever structure remains in the CM diagrams after
the removal of this disk structure is attributable to the bulge. We show
in figure \ref{ratios} a quantitative  comparison of the model for
latitude $b=0.5$\degree and the  observations (figures \ref{model0.5} and
\ref{b0.4-0.6}). The figure shows  the ratio of sources observed to those
predicted in the model as a contour map over the (J$-$K)-(I$-$J), (J$-$K)-(K)
and (I$-$J)-(J) planes. The dotted contours indicate small variations
likely due to patchy extinction, for which account cannot be made in this
model. The three dotted contours represent  ratios of 0.5, 1.0 and 2.0.
The solid contours represent number ratios from 4  up to 12. It is clear
that in the (J)-(I$-$J) diagram, there is very little  difference between
the model and the observations. As expected for a diagram limited by the I
band, distance penetration is low and only the disk is seen. However, for
the (J$-$K)-(K) diagram, there is clearly an excess population of sources
peaking at a colour of J$-$K$\sim$3.5. This may be identified with a
reddened giant branch in the bulge, which is in accord with the
expectation that the power of the J and K bands to penetrate dust will
allow the bulge to be seen.  There remains a hint of this effect in the
reddest sources of the two-colour diagram (J$-$K)-(I$-$J), but as for the
(J)-(I$-$J) diagram, this is limited by the limited distance penetration of
the I band, and will be well represented by a disk only. The ratio of
bulge to disk sources seen here at $b=0.5$\degree of \alt 12 is in good
agreement with the result (\alt 15) we derive in a later section from
published bulge/disk ratios.

The derivation of the parameters for the exponential dust-disk is detailed in
the next section.

\begin{figure}
\label{ratios}
\end{figure}

\section{Deriving the dust parameters}

\subsection{Elementary characterisation of dust layer}

It is clear by looking even at integrated light images of the Galactic
Central regions (e.g. Madsen et al., 1986) that there is a prominent
dark band in the plane of the galaxy. This is due to the dust which
pervades the disk, and causes light to be attenuated, especially at
short wavelengths. The feature is  diminished at longer wavelengths.

If we consider the dust to be a uniform plane of finite thickness
$2d$, extending in the direction of the plane of the galaxy, with the
sun centrally placed, we can easily calculate that the path length in
the dust layer is given, in terms of the galactic latitude, $b$, by
$d\ cosec\ b$. This optical path retains the same functional
dependence on $b$ if the uniform layer is replaced by a dust
distribution with an exponential dependence (For a vertical
scaleheight of $z_h$, the result becomes just $z_h\ cosec\ b$).

Figure \ref{colourbcut} shows the mean colours for cuts of height
0.2\degree\  covering the full range of the dataset from 5\degree\ to
$-$5\degree. The colour shown is found by taking the mean value of all
sources in the relevant colour-magnitude diagrams.

The variation with mean colour with latitude is mainly symptomatic, at
these low latitudes, of extinction in the plane.

\begin{figure}
\label{colourbcut}
\end{figure}

The major features in figure \ref{colourbcut} are the clear rise in
mean J$-$K colour towards the plane, and the equally clear fall in mean
I$-$J colour towards the plane. These can both be understood in terms of
different degrees of reddening and different populations sampled.

For the J$-$K diagram, the penetration into the dust is limited by the J
waveband completeness.  The dominant sources, seen to large distances,
are type III giants, and the closer the approach to the plane, the
greater the amount of dust in the line of sight, leading to a peaking
in the mean colour towards the plane.

For the I$-$J diagram, the same effect will clearly be taking place, but
the limitation in this figure is the effect of the dust convolved with
completeness in the I waveband.  Towards the plane, the distance
observed is low, and the sources are dominated by nearby,
main-sequence, blue stars. These have an intrinsic colour from
I$-$J=$-$0.2 to 0.5. As we look away from the plane, due to the decreased
dust in the line of sight, we see to greater distances, and sample the
giant stars visible to a much larger distance. The figure is
effectively the same as the J$-$K figure (as suggested by the tails),
with a large wedge removed from the middle due to the limited distance
penetration at the wavelength of the I band.

It is also clear in the figure that there is an asymmetry about b=0
for the mean J$-$K colour of extracted sources. The peak, in fact,
appears at b$\sim$0.15\degree. This can be attributed to a non-zero
height for the sun above the plane of the local dust.

We can thus, using the I$-$J diagram, place a very approximate limit on
the height of the extinction layer by noting that the abrupt change
occurs at $\sim\pm1$\degree, and that the brightest main-sequence stars
may are visible to distances of 2--3\,kpc \cite{n+k80}. One degree at
this distance corresponds to about 40--50\,pc, and can be seen as an
indication of the scaleheight of the dust.

Similarly, a simple estimate of the displacement of the sun can be
made by noting that in J and K bands, the typical colour of objects
seen to any significant distance in the (dusty) plane is between 0.6
and 1.0 (type III objects -- see figure \ref{cmstandard}).

Indicated on the left in figure \ref{colourbcut} are $cosec$ law fits
of the tails of the distribution of extinction, using an intrinsic
colour of J$-$K=1.0.  The fits correspond to scaleheights differing by
about 40\%. i.e.

 $$  \frac{z_d-z_0}{z_d+z_0} = 1.4 $$

where $z_d$ is the scale height of the dust, and $z_0$ is the
displacement of the sun above the plane. The solution we obtain is
that $z_0 / z_d \sim -0.17$.  The fit is very insensitive to the
intrinsic colour adopted. (e.g. changing (J$-$K)$_0$ from 1.0 to 0.6
changes $z_0 / z_d$ from $-$0.17 to $-$0.15).

Using the scale height of about 40--50\,pc estimated above, we deduce
that the sun lies about 7\,pc below the local galactic dust plane.

And finally, noting that the reddest J$-$K colour of sources near b=0 is
some 2.0--2.4 magnitudes redder than the intrinsic colour
(corresponding to an $A_V$ of 18--22), we can estimate, assuming DENIS
sees some giants at the Galactic centre at about 8\,kpc distance, that
$A_V$ is on average roughly 2.5 mag/kpc.

These estimates for the scale height, distance from plane, and
$A_V$/kpc value can be refined in a model-dependent way.

\section{Model dependent derivation}

The dust layer, in this simple model, is characterised by three
parameters : the scaleheight of the assumed exponential profile
($z_d$), the radial scalelength ($r_d$) and the local rate of
extinction ($A_V$ in mag/kpc). In addition, the displacement of the
sun from this midplane must be included ($z_0$).

The rough method described above provided initial estimates for the
values of these three parameters -- $z_d$,$A_v$ in mag/kpc and $z_0$.
For the fourth parameter $r_d$, we use a value of 2.7\,kpc as for the
stellar disk, as a starting guess.

The model described above was set up with the adopted parameters for
the stellar disk and dust scalelength, and the three parameters to be
determined by a fit to the data (i.e. $r_d$,$z_d$,$A_v$ in mag/kpc and
$z_0$).

A 5$\times$5$\times$5$\times$5 grid of synthetic colour-latitude
diagrams was made by generating stars according to the model described
above, using all the possible combinations of the following values of
the three parameters:

\begin{itemize}

\item $r_d$=2,4,6,8,10 kpc

\item $z_d$=20,30,40,50,60 pc

\item $z_0$=$-$25,$-$20,$-$15,$-$10,$-$5 pc

\item $A_V$ = 0.5,1.0,1.5,2.0,2.5 mag/kpc

\end{itemize}

The resulting point source data were convolved with the
two-dimensional completeness functions derived in section 3
for the I$-$J and J$-$K data, and  were subsequently used to generate
colour-latitude diagrams by linear interpolation between points in
this data tesseract. The statistic minimised is the sum of the squares
of the difference between observed and model colours summed over both
I$-$J and J$-$K colours. The ranges in latitude in each case were limited
by practical factors.

In J$-$K, the ranges for comparison were limited to the tails of the
distribution beyond J$-$K=1, so as to avoid the strongest bulge
contamination. The tails of the J$-$K distribution are  expected to
sample to large distances, thereby averaging out, to some extent,
local inhomogeneities in the dust distribution.

The same cannot be true for I$-$J - this will be severely affected by
local inhomogeneities due to the limited distance sampled by the I
waveband. Contrary to the situation in the J$-$K figure, we expect the
lowest latitudes to be little, if at all, contaminated by bulge
sources, and a fit there is appropriate. We do, however, exclude the
highest latitudes (b$>$1) due to the known presence of very nearby
(few 100\,pc) dusty star forming complexes at positive latitudes which
may severely bias the result (e.g. $\rho$ Oph)

The 18 points thus chosen for the fit are the data points at latitudes
$b=-$1.2,$-$1.0,1.0,1.2,1.4,1.6,1.8\degree\ in J$-$K, and all points between
$-$1.2\degree\ and 1.0\degree\ inclusive in I$-$J. All were given equal weight
in a least squares fit. The statistic $X$ used was simply the sum of
squares of differences between the model and the observations:

i.e. \[ X = \sum_{i=1}^{N} (c_i - c(b_i;a_1,a_2 ... a_M))^2 \] where
$c_i$ denote the observed mean colour values and $c(b_i;a_1,a_2
... a_M)$ denote the model derived mean colours. $b_i$ is the latitude
associated with that colour point, and $a_1 ... a_M$ are the $M$
parameters associated with the model. In the present case, M=4, and
$a_1$,$a_2$,$a_3$ and $a_4$ are local $A_V$ in mag/kpc,$z_d$,$r_d$
and $z_0$.

The number of data points, N, is 18 since the summation is over the
limited set of 18 points in the I$-$J colours and the J$-$K colours from
$-$1.2\degree\ to 1.8\degree\ inclusive in 0.2\degree\ steps.

This statistic can be used to estimate the standard deviation associated with
each data point by means of  the following formula \cite{press92}:

\[ \sigma^2 = \frac{\sum_{i=1}^{N} (c_i - c(b_i;a_1,a_2 ... a_M))^2}{N-M} \]

Minimising the statistic X with respect to variations in $A_V$,
$z_d$,$r_d$ and $z_0$, we find a value X=0.484 occurring at values
$A_V$/kpc = 1.40, $z_d$ = 40\,pc, $r_d$ = 3.4\,kpc and  $z_0$ = $-$14.0\,pc.
An estimate for $\sigma$ is thus 0.13, which agrees well with
inspection of the plots. A value can thus be assigned to $\chi^2$ by
using the standard formulation:

\[ \chi^2 = \sum_{i=1}^{N} \left(\frac{c_i - c(b_i;a_1,a_2 ... a_M)}{\sigma}\right)^2 \]

Subsequently, a 1$\sigma$ estimate of the uncertainties in the
parameters derived can be obtained by looking for the variations in
those parameters which give rise to an increase in $\chi^2$ of
$\Delta\chi^2$ = 1.00 \cite{press92}. This corresponds to a
value $\Delta X$ of 0.0346.

The result obtained in this way is:

\begin{itemize}

\item $A_V$/kpc = 1.40 $\pm$ 0.11

\item $r_d$ = 3.4 $\pm$ 1.0\,kpc

\item $z_d$ = 40 $\pm$ 5\,pc

\item $z_0$ = $-$14.0 $\pm$ 2.5\,pc

\end{itemize}

The model and observations are compared in figure \ref{collatfit}. The
fit to the wings of the J$-$K distribution is very good, which is as
expected since the J$-$K sample is expected to sample a large path
length of the disk, and any small scale inhomogeneities, such as the
conspicuous +ve latitude nearby dust regions, are averaged out. The
fit to the  I$-$J data is less convincing as local structure in the
interstellar dust strongly biasses the mean colours.

\begin{figure}
\label{collatfit}
\end{figure}

Note that the above method does not use a truly independent value for
$\sigma$.  This is something which cannot be readily defined for the
dataset. The uncertainties are not measurement error, but are due to
the random nature of the extinction in the line of sight. As a
consequence, the uncertainties in the results above reflect the
uncertainties of this best fit within these limitations. They do not
in any way indicate to what extent the functional forms are justified.

These dust distribution parameters are used in the model of the galaxy
described above.

\subsection{Comparison to other results}

A whole sky fit to the COBE/DIRBE data has been made by Freudenreich
(1996). He excluded the difficult central region ($\sim$ 40\degree\
$\times$ 30\degree) as well as other parts of the plane, and fitted a 28
parameter model, constraining simultaneously the scales and intensity
of the dust layer, and the stellar disk.

Converting his dust layer parameters to the units used here, he gives:

\begin{itemize}

\item $A_V$/kpc = 1.53$\pm$0.01

\item $r_d$ = 3.85 $\pm$ 0.10\,kpc

\item $z_d$ = 46$\pm$1\,pc

\item $z_0$ = 15.55$\pm$0.23\,pc

\end{itemize}

Freudenreich uses a functional form sech$^2(z/2z_d)$, as opposed to the form
$\exp(-z/z_d)$ used here. These two forms are equivalent for $z \gg z_d$
(within a numerical factor of 4), but importantly, near $z=0$ in the
plane, these functions differ significantly. At height zero, they differ by a
factor of 4, and at heights of $z_d$ and 2$z_d$, still by factors of 1.8 and
1.3. The fits made by Freudenreich exclude regions within a few degrees of the
plane in most places, and he chooses the sech$^2$ functional form because it
has some basis in theory as the density law of an isothermal self-gravitating
disk \cite{k+s81}. We find here roughly the same value of the parameter for
scaleheight, but we assume an exponential form for the dust layer.

The values obtained for the intensity of local extinction, and the
scale length of the dust layer are in good agreement. Integrating this
model in a line of sight towards the centre results in a value for
total $A_V$ of 56. The wide uncertainty in $r_d$ results in large
uncertainties on this value between 40 and 110. Many observations
towards the galactic centre (e.g. Becklin \& Neugebauer 1968; Catchpole, Whitelock
\& Glass 1990) seem to agree
on a value of extinction in the line of sight towards the very centre
of the galaxy as $A_V \sim 40$, and our results are consistent with
this.

The value we find for the distance of the sun from the local dust plane
suggests that it lies below the plane of this dust (i.e. towards the South
Galactic Pole). Indeed, looking at the surface photometry map produced using
DIRBE data at 1.25$\mu$m (Weiland et al. 1994 -- their figure 1), we see that
the part of the bulge at negative latitudes appears brighter than that at
positive latitude. Taken at face value, this would suggest a greater path
length of dust towards the northern part of the bulge, and hence a position
below the dust plane. However, as is clear from the same map after correction
for extinction (Weiland et al. 1994), the most appropriate position for the sun
is above the plane of the disk. This is further corroborated by many other
studies using widely different methods. For example, Cohen (1995) finds a
distance of 15\,pc above the plane by comparing north and south Galactic pole
star counts. Figure 3b of Freudenreich et al. (1994) shows that the DIRBE  240$\mu$m
emission lies below the Galactic plane, and though the relation between
240$\mu$m flux and dust is not well calibrated, this nevertheless suggests a
displacement above the plane for the sun.
Binney, Gerhard \& Spergel\footnote{Note that the preprint version
(Binney, Gerhard \& Spergel 1996) and
the published version of the article by  Binney, Gerhard \& Spergel (1997)
differ in the position ascribed to the sun. The preprint consistently states a
position of 14\,pc {\em below} the disk plane for the sun, while the published
version states a value of 14\,pc {\em above} the plane.}
(1997) also find a value of 14\,pc above the plane after modelling the inner
Galaxy. Several more studies (e.g. Conti \& Vacca 1990; Toller 1990) all agree
on a value of $\sim$10--20\,pc above the plane. Clearly, all work to date is in
agreement that the sun lies above both the Galactic stellar and dust planes.

Our result is not sensitive to the stellar plane, but rather to the dust plane
-- the stellar plane acts only as a luminosity source of near-constant colour.
The simple model we construct imposes a global symmetry, and is fitted based on
data from a very limited set of directions, unlike most of  the other
references cited above. We did not expect to derive globally reliable
parameters which describe the three-dimensional complexity of dust distribution
present in the disk/bulge, and indeed the position we derive for the sun is a
demonstration of this fact. There is clearly asymmetry in the local extinction
(for example, there is substantially greater extinction present north of the
Galactic plane, much associated with the nearby $\rho$\,Oph star-forming
region, than to the south). Presumably it is this which the analysis of the
DENIS data is sensitive.

\section{Peculiarities in the Colour-Magnitude diagrams}

Some of the peculiarities described above can be understood in a model
dependent way.

\subsection{Asymmetric Scatter to the right of the two colour line}

Scatter can be seen to the right of the two-colour line as in figures
\ref{blt0.1},\ref{b0.4-0.6},\ref{b1.0-1.2}. If real, these would
represent objects with very high intrinsic J$-$K colour, compared to I$-$J
colour. Figure \ref{cmstandard} indicates that normal stars do not
appear like this. It is possible that some of these are dust-enshrouded stars,
which shine brightly at longer wavelengths, but are very much more
obscured at shorter wavelengths. But before turning to these
astrophysical explanations, we shall consider the data reduction.

\begin{figure}
\label{twocol1}
\end{figure}

Figure \ref{twocol1} shows the two-colour diagram for a typical
crowded field, when the matching radius, within which all three images
(I,J and K) must fall in order to be accepted, is varied between 0.3$"$
and 3$"$. It is clear that as the radius is decreased, spurious points
scattered to the right of the line become fewer.
If the matching radius for finding common
sources between the 3 images is left too large (e.g. 3$"$), there are
many sources apparently very far from the line. Reducing the matching
radius to 1$"$ loses some 10\% of sources of deviation zero, but loses
over 80\% of those with deviation 2 magnitudes.

\subsection{Bent two-colour line}

In looking at the highest latitude colour-colour diagrams, (such as
figure \ref{b1.0-1.2}), we notice that I$-$J colours are too red
compared with J$-$K colours at the same reddening. The two-colour line
is effectively bent upwards.

AGB stars are expected to be found in this region of the colour-colour diagram.
(Groenewegen, 1997) But another artifact of data processing can explain at least
part of these bends, by attributing them systematic errors in photometry caused
by crowded fields. The effect is well reproduced in the model when the
systematic offsets in magnitude are included. Figure \ref{model1.1} shows a
clear upward bend in the reddest parts of the two colour diagram. This can be
understood in terms of the systematic effects in magnitudes. In particular, the
J-band magnitudes show a very large offset for the faintest magnitudes
retained, larger than those of the corresponding $I$ and $K$ magnitudes. The
offset, for the reasons described above, causes the magnitudes to be too bright,
or numerically, too low. This means that J$-$K is too low, and I$-$J is too
high. This is precisely the effect seen in both the simulation and the
observations.

\subsection{Broken striations}

Figure \ref{c02} shows a set of diagrams where striations in the
colour-magnitude diagrams are fractured in the direction of the reddening
line. This too is well understood in terms of the model by allowing a
`wall' of extinction in the line of sight, as demonstrated by figure \ref{modelwall}.

\section{Distance distributions}

The value of the simulated datasets, which are a good match to the
observations, is that they can be interrogated for distance
information, which will help to deduce the three dimensional
distribution of stars in the inner galaxy. The information obtained
in this way is summarised for the simulation at $b=0.5$\degr in
figure \ref{dist}. This particular latitude has been chosen to enable
the dust to be used to our advantage to separate source in the near
disk and in the inner disk and bulge. At higher latitudes, the
extinction due to dust is much less and penetration is good in all
three bands. Near and far sources are not spread out by the reddening
in the line of sight to a great extent. On the contrary, at very low
latitudes approaching zero, the extinction is so severe that, given
the magnitude limits of the system, it is not possible to see past
the nearby disk population into the bulge regions, except possibly at
K band.

\begin{figure}
\label{dist}
\end{figure}

From figure \ref{dist}, it is clear that in the I band, due to the
inpenetrability of the local dust at short wavelengths, even the
faintest magnitudes fail to penetrate very far from the sun.  The
situation changes at J band, where while penetration is low (\alt 8\,kpc)
at the lowest magnitudes (J=9.0--11.0), there is some penetration
of the bulge regions at the fainter magnitudes (J=13.0--15.0). At the K
band, the penetration is even greater. Even the very brightest
magnitudes may penetrate to the bulge region, but at the fainter
magnitudes, the source counts are dominated by stars near the
bulge. Remember that this simulation includes only a disk component,
so when the model suggests penetration in as far as the bulge, the
counts at that magnitude are likely to be dominated by bulge objects.
This is shown in the contour plot above (figure \ref{ratios}), where
the ratios of bulge to disk objects will be at \alt 12.

\section{Disk/Bulge asymmetries}

\subsection{Bar models}

There has been recent interest in the possibility of a kiloparsec
scale bar  at the centre of our galaxy. Various lines of evidence
suggest the presence of a bar, and though they disagree on the exact
parameters which best describe the form of the bar, they agree that
the major axis is oriented  towards the first quadrant. Methods
employed include gas dynamics (Binney et al., 1991; Blitz \& Spergel
1991), and modelling of integrated light distributions from the
COBE/DIRBE experiment (Dwek et al., 1995; Binney et al., 1997). The
asymmetries expected in projection are such that number counts  at
equally positive and negative longitudes should in general be greater
at  {\em positive} longitudes for large longitudes, and at {\em
negative} longitudes at small longitudes.

Integration along the line of sight in such models for equal positive
and negative longitude pairs results in different number counts for
the two lines of sight. In figure \ref{barassyms}, the number count
asymmetries predicted by various models are shown. The effect of
luminosity function will be small at these longitudes. The difference
in distance modulus to the main concentration of the bar will be \alt
1.0 magnitude, since the separation in angle is at most 10\degree.

In practice, the amplitudes shown here will only be realised when a
tracer population which samples only the bulge is used. In the
present case, there is much disk contamination, and we expect, in
general, a lower signal to be seen. An estimation of the extent of
this dilution is given by integrating a recent model fit to the
bulge/bar and the disk of the galaxy by Binney et al. (1997). For the
purposes of this discussion, the contrast between disk and bulge in
the inner disk is important. Using the model there, we find that for
sources inward of 3\,kpc from the centre, observed from the sun, and
for longitudes of \alt5\degree,  the number count contrasts vary
strongly only with latitude (because of the thinness of the disk).

\begin{tabular}{ccccccc}
 b(deg)     & 0.0 & 0.2 & 0.4 & 0.6 & 0.8 & 1.0 \\
f$_{disk}$/total & 0.12 & 0.08 & 0.06 & 0.05 & 0.04 & 0.03
\end{tabular}

These ratios are small, and in the light of other uncertainties
present in this analysis, may be neglected.

\begin{figure}
\label{barassyms}
\end{figure}

In order to optimise detection of the inner-disk/bulge asymmetries in
our data, we can use the information furnished by figure
\ref{dist}.
We then test this for left-right asymmetries in the
disk/bulge at various distances. We must first assume the similarity
of disk luminosity functions at equal positive and negative longitudes
for a give latitude of observation. This is not a contentious
assertion, as the difference in the lines of sight differs, in all the
following cases, by less than 10\degree, so systematic age or
metallicity differences are not expected.

The effect still remaining in the data which prevents immediate
comparison between number counts at equal and opposite longitudes is
caused by patchiness in the extinction, as there is no reason to
expect this to be systematic.  To minimise this effect, we use the
model to identify magnitude ranges dominated by disk objects, and
identify and correct for any associated (foreground) asymmetry.

That is, cuts in magnitude are chosen which are dominated by disk
objects. A fit is made at these magnitudes between equal negative and
positive longitude pairs. Any additional asymmetries remaining at
fainter magnitudes will (in the case of the J band and K band)
contain  some signal of asymmetries in the inner disk or in the
bulge. The I band, according to the model distances, should serve as
a control, since it is not expected to penetrate very far, and the
number counts seen should, after this correction for differences in
extinction, show equal values at positive and negative longitudes.

\section{Results of this experiment}

\begin{figure}
\label{matchsym}
\end{figure}

The magnitude limits chosen for the cuts are as follows:

\bigskip

\begin{tabular}{ccc}

Band & Fit region & Test region \\
     &  ``Disk''  &  ``Bulge''  \\
\\
I & 11.0--14.5 & 14.5--17.0 \\
J & 9.0--11.0 & 11.0--13.0 \\
K & 7.5--9.0 & 9.0--10.5
\end{tabular}

\bigskip

Figure \ref{matchsym} shows the results for $b=0$ when cuts are made
as described above to match the local disk luminosity function. The
contrasts shown are for the fainter magnitudes as indicated above.

\subsection{Deviation from unity}

Do the contrasts deviate, in the mean, from unity? Taking the mean
values of the contrasts in the three cases and finding the deviation
of the mean we find the following:

\bigskip

\begin{tabular}{cccc}

Band & Mean, $\mu$ & $\sigma_{\mu}$ & ($\mu$-1)/$\sigma_{\mu}$ \\
I & 0.961 & 0.026 & $-$1.5 \\
J & 1.105 & 0.037 & 2.8 \\
K & 1.118 & 0.039 & 3.1
\end{tabular}

\bigskip

The suggestion is that the I band deviates insignificantly (no more
than 1.5$\sigma$) from unity. This is as expected, since the I band
counts, though patchy, once matched for a pair of directions according
to the local distribution of the brightest sources, show no further
differences at faint magnitudes. The distance model above suggests
that the penetration is not nearly deep enough into the disk to allow
the innermost parts of the disk, or the bulge, to be sampled.

The J band shows a more significant deviation from unity in the ratio
of counts ($\sim$ 2.8$\sigma$). The distance model in this case
suggests that penetration is deeper, and may reach the inner disk and
bulge. Indeed, the contrasts seen suggest that counts at negative
longitudes are greater than those at corresponding positive
longitudes. Similarly, in the case of the K-band number counts, we
see a similar asymmetry. In this case, there is a marked dip in the
contrast at a longitude of about 2 degrees. This is most plausibly
due to a mixture of structural and extinction effects close to the
centre of the galaxy. In the analysis here, we try to remove
asymmetries in the nearby disk caused by dust to show up asymmetries
in the inner disk or bulge. It is clear from the $^{12}$CO($^1 J
\rightarrow ^0 J$) intensity plot shown in figure \ref{littlecomap}
\cite{dame87} that there are distributions of material towards the
centre of the Galaxy with intensities differing by several orders of
magnitude over a very few degrees. In this light it is not
surprising that the contrasts seen in the K number counts would
not show any clear bar-like signature even if one existed. One
approach to combatting this problem is to obtain multicolour
information for sources in these regions and deredden each on a point
by point basis. This technique is used in paper 2 to treat K and L
band data.

\begin{figure}
\label{littlecomap}
\end{figure}

The results here are at best inconclusive. The contrast seen in the K band
is always, in the mean, greater than unity, suggesting that the
asymmetric inner disk dust effects traced in figure \ref{littlecomap}
are a perturbation on a net greater negative longitude count compared
with positive longitude counts.

A similar effect is not seen clearly in the J-band, which suggests that
the realm of this central asymmetry, be it structural or due to dust, is
not reached at this shorter wavelength.

\section{Conclusions}

We have derived some techniques to extract photometric and astrometric
information from DENIS images in crowded fields and to characterise its
deficiencies. We have used this dataset, covering a part of the region
within $|l|<$5\degree, $|b|<$1\degree, to construct a model of the
Galactic disk, and to fit parameters for a model of the dust layer.
Using this model, we find that the large numbers of very red sources in
the observed colour-magnitude diagrams, not reproduced by the model,  can
be understood as bulge giant branch stars. We assume the symmetry of the
structure and luminosity function in the disk in directions of equal and
opposite longitudes and fit number counts for a bright cut in magnitude
(corresponding, according to the model, to near disk sources), and look
for asymmetries in the fainter number counts (corresponding to inner disk
and bulge objects). We find that in the I band, there is no asymmetry at
fainter counts, consistent with the model-based expectation that the I
band does not penetrate very far into the disk. In J and K bands, there
is $\sim$3$\sigma$ evidence for a ratio of negative longitude to positive
longitude  number count which is greater than one. This is consistent
with the expectation from bar models.

This large scale statistical approach allows, to some extent, the
possibility of `averaging out' localised anomalies in extinction which
are common in the plane, and towards the centre of the galaxy. However, as
can be seen especially in the K-band contrasts, number counts can be
seriously affected by the distribution of dust regions near the centre of
the Galaxy itself. In the present set of observations, it is not clear
that the J-band data penetrates the densest dust regions reliably. One
possibility is to use longer wavelengths where the extinction coefficient $A_X$/$A_V$ is lower
still, so that dust is a less severe problem. This method is explored in
paper 2, where we combine DENIS K data and UKIRT nbL data (3.6\mum,
$A_{nbL}$ = 0.047 $A_V$).

\section{Acknowledgements}

The DENIS project is supported by the {\it SCIENCE  and the Human Capital and
Mobility }  plans of the European Commission under grants   CT920791 and
CT940627, the European Southern Observatory, in France by the {\it Institut
National des Sciences de l'Univers}, the Education Ministry and the  {\it
Centre National de la Recherche Scientifique}, in Germany by  the State of
Baden--Wurttemberg, in Spain by the DGICYT, in Italy by the Consiglio Nazionale
delle Ricerche, in Austria by the Science Fund (P8700-PHY, P10036-PHY) and
Federal Ministry of Science, Transport and the Arts, in Brazil by the Fundation
for the development of Scientific Research of the State of S\~ao Paulo (FAPESP).

MU would like to thank GS at the Observatoire de Paris for his hospitality
during visits there, as well as Francine Tanguy, Jean Borsenberger and Lionel
Provost for their help with access to the DENIS archive. MU acknowledges the
financial support of the Particle Physics and Astronomy Research Council.

\section{Appendix -- The Galaxy model calculations}

A Monte-Carlo method for generating stars in a Galactic model is mentioned
above. Details of the method employed for optimising the Monte-Carlo process
are given below.

For a given line of sight, a probability distribution is generated, $p(r)$,
corresponding to the number of sources seen, per unit distance, due to the
geometrical components included in the model. Clearly, because of the central
concentration of the galaxy, this will mean that the peak in this distribution
will lie at approximately $r_0$ from the sun towards the Galactic Centre (GC),
where $r_0$ is the sun-GC distance.

Also for this line of sight, based on the dust model, a function $A_V(r)$ can
be generated, which gives the extinction to any distance in that line of
sight.

The luminosity function (LF) used has logarithmically more faint sources than
bright sources.

Thus, if, independently, a random position is chosen according to $p(r)$ and a
random stellar type is chosen according to the LF, the resulting pair will most
likely be a faint, main sequence star near the Galactic Centre -- which will not
be visible.

The Monte Carlo approach used to generate random points in this two
dimensional space of distance and luminosity function will in general be very
wasteful. We derive a method for limiting the
space in which random points are thrown, and which also ensures
that the very low probability near and faint objects are accurately included in
the number counts.

The functions embodied in the model generate, for a given line of sight, the
PDF (the probability per unit distance that a point will be found there) and
$A_V(r)$ curves (integrated flux diminution up to that distance). Examples
are shown for a line of sight in the direction $\ell=$4.0\degree\ and b=0.2\degree\
in figure \ref{stages1}.

\begin{figure}
\label{stages1}
\end{figure}

\subsection{Distance}

First of all, the function $p(r)$ is converted to a function of distance
modulus, $D_m$. Noting that

\[ p(r)dr = p(D_m)dD_m \] and that
\[ D_m = 5\lg(r/kpc) + 10 \]

we obtain

\[ p(D_m) = \frac{dr}{dD_m} p(r) \propto rp(r) \]

This function is then numerically integrated to give a function

\[ F(D_m) = \frac{\int_{-\infty}^{D_m} p(D_m')dD_m'}{Q} \]

where the normalisation $Q$ is given, in terms of the upper limit $D_m^0$ in
$D_m$  chosen for the simulation, by:

\[ Q = \int_{-\infty}^{D_m^0} p(D_m')dD_m' \]

Figure \ref{stages2} shows both $p(D_m)$ and $F(D_m)$ for the example shown in
figure \ref{stages1}.

\begin{figure}
\label{stages2}
\end{figure}

This function $F(D_m)$ is subsequently numerically inverted, so that the input
of a random deviate uniformly distributed between 0 and 1 will result in an
output $D_m$ distributed according to the PDF for the line of sight.

\begin{figure}
\label{stages3}
\end{figure}

Figure \ref{stages3} shows the result of this inversion, with the ordinate
$\ln x$, where $x$ is a uniformly distributed deviate between 0 and 1.
Furthermore, corresponding to each distance modulus, $D_m$, we can assign a
value $A_V$ from the lookup diagram shown in the upper panel of figure
\ref{stages1}. Each waveband, I,J and K, corresponds to a different
value of extinction, $A_I$,$A_J$,$A_K$. These three modified distance
modulus curves are shown in the figure.

\subsection{Luminosity Function}

The luminosity functions may be treated in a similar way to the distance
modulus distribution. Taking the K-band luminosity function as an example, we
can, as before, generate a cumulative distribution, invert it, and use this as
a lookup table to convert a deviate, $y$, uniformly distributed between 0 and
1 into a corresponding stellar type.

\begin{figure}
\label{stages4}
\end{figure}

\subsection{Combination}

To limit the numbers of sources which are tried only to those brighter than
some given magnitude limit, let us denote by $x$ and $y$ two independent
uniform random deviates between 0 and 1. From $x$, we derive a modified
distance modulus according to the distance modulus and reddening combined
lookup table, as in figure \ref{stages3}. From $y$, we deduce a stellar type
in the same way. The overwhelming likelihood is that a faint source at large
distance is generated, which will clearly fall out of the magnitude range of
interest.

In figure \ref{busyplot}, there are a multitude of lines. Concentrating on the
long-dashed line, we see that combining the $x$ figure for modified distance
modulus for K-band, and the $y$ figure for K-band luminosity function, limits in
observed K magnitude are expressed by lines roughly parallel to lines of
constant  $w = \ln x + \ln y$, or constant $xy$. Similarly, magnitude limits can
be set for J and I, leading to the limiting lines shown  (dotted and dashed
lines).

\begin{figure}
\label{busyplot}
\end{figure}

\subsection{Generation}

The aim is to take two independent uniform deviates between 0 and 1, $u$ and
$v$, and convert them to deviates $x$ and $y$ which uniformly cover the space
$(x = 0 \rightarrow 1, y=0 \rightarrow 1)$ {\em excluding} the region with
$w<w_0$.

Graphically, we seek to fill only the left region in figure \ref{appfigure},
where the bounding function is $xy = e^{w_0}$.

\begin{figure}
\label{appfigure}
\end{figure}

Now for uniform deviates $x$ and $y$, the product $z = xy$ will be distributed as

\[  p(z)dz = \int_{z}^{1} dx \int_{\frac{z}{x}}^{1} dy -
             \int_{z+dz}^{1} dx \int_{\frac{z+dz}{x}}^{1} dy \]

or

\[ p(z) = -\ln z \]

To well sample the lowest probabilities, we convert this to a function in terms
of $w = \ln z$, to give $p(w) \propto -we^w $. We then integrate and
normalize this function to find a cumulative form, with a lower limit
$w_0$ and upper limit  $w_1$ ($w_0$,$w_1 \le$ 0). The result is:

\[ F(w) = \frac{ e^{w_0}(1-w_0) - e^w(1-w)}{ e^{w_0}(1-w_0) - e^{w_1}(1-w_1)} \]

As before, when this is inverted, a uniform deviate, $u$ between 0 and 1 can
be supplied, and the value $w$ corresponding to $F(w)=u$ gives a value for
$w = \ln x + \ln y$ such that the points $(x,y)$ are uniformly distributed in
the allowed region.

By symmetry, $x$ and $y$ are distributed in the same way for any given value
of $xy$. We can thus find values $x$ and $y$ by using the second uniform
deviate $v$. Constructing $d = w(2v-1)$, we obtain a uniform deviate between
$-w$ and $w$. This corresponds to the difference $\ln
x - \ln y$. This leads to the final result, that if a deviate $w$ and a
secondary deviate $d$ are generated as described  above, then

\[ \ln x = \frac{d+w}{2} \]

and

\[ \ln y = \frac{d-w}{2} \]

where points $(x,y)$ are uniformly distributed in the allowed region.

The amount of the $x-y$ plane excluded by limiting sources to the left of the
line $xy=z_0$ is given by $z_0(1-\ln z_0)$, or $e^{w_0}(1-w_0)$ where $w_0
= \ln z_0$.

In the example illustrated, a limit of $w_0$=$-$10 allows all observable
sources to be generated. This corresponds to sampling only 1/2000th of
the $xy$ plane.  In general, we find that $w_0$ lies between $-$7 and
$-$10, allowing savings in processor time by a factor of between 140 and
2000. The algorithm described was translated into the Super Mongo (SM)
programming language before use.

\subsection{Comments about Spiral Arms}

It is not
clear what the contrast between arm and interarm regions is likely to
be at near IR wavelengths. The dominant emission is from giant stars,
which will plausibly have diffused away from the sites of young star
formation associated with spiral arms. Furthermore, dust associated
with the spiral arms, which plays a role in increasing the contrast at
short wavelengths, has a lesser effect at longer wavelengths. Hence
the contrast is likely to be less beyond 1\mum\ than below it. Studies
of external disk galaxies \cite{jong96} suggest that the contrast
may be \alt 3, which is in agreement with recent DIRBE based
model-dependent analysis of our galaxy \cite{bgs97}. It is not even
certain that well-defined spiral arms are observable at all
wavelengths. Independent observations using bright O stars to probe
the disk \cite{n+k80} and HI regions \cite{g+g76} yield different
spiral arm structure.

However, the limited directions to which the model is to be applied in
this case -- the inner 10\degree\ of the galaxy -- means that lines of
sight will in general pass perpendicularly to the directions of any
spiral arms which may be in  the disk. For the purposes of
investigating longitude dependent asymmetries, lines of sight
perpendicular to the arms and separated by no
more than 10\degree\ will not have a very different amount of spiral arm
signature present.

For these reasons, and anticipating the method used to remove near disk
signature from number counts, we do not include spiral structure in the
model. For reference, a sample set of colour-magnitude diagrams is shown
when spiral arms are included.  Spiral arms of gaussian width 0.5\,kpc are
placed in the line of sight at distances 3.4, 5.1 and 6.9\,kpc from the
Galactic centre in accordance with the model of Georgelin \& Georgelin
(1976), and are given an enhancement of a factor of 3 with respect to the
underlying disk density. (figure \ref{spiralincluded}). No modification
is made to the dust model. There is only a slight hint of difference
between this model and the comparison model without spiral arms (figure
\ref{model0.0}), and it is not useful at this level, to include this
subtlety in our model.

\begin{figure}
\label{spiralincluded}
\end{figure}

\end{document}